\documentclass{ws-ijmpd}

\usepackage[super]{cite}      
\usepackage[dvipsnames]{xcolor}

\usepackage{amsfonts}
\usepackage{amsmath}

\usepackage{url}
\usepackage{circledsteps}
\usepackage{graphicx}
\usepackage[misc]{ifsym}
\usepackage[caption=false]{subfig}

\usepackage{microtype,orcidlink}

\usepackage{hyperref}
\hypersetup{verbose,hypertexnames=false,colorlinks=false,allbordercolors=Blue,pdfborderstyle={/S/U/W 1}}

\begin{document}

\markboth{D.\ G.\ C.\ McKeon, F.\ T.\ Brandt, J.\ Frenkel and S.\ Martins-filho}
{A Renormalizable and Unitary Approach to Quantum Gravity}

%
\catchline{}{}{}{}{}
%

\title{A Renormalizable and Unitary Approach to Quantum Gravity\footnote{This essay received an Honorable Mention in the 2026 Essay Competition of the Gravity Research Foundation.}}

\author{D.\,G.\,C.~McKeon}

\address{
   Department of Applied Mathematics,\\ The University of Western Ontario,\\ London, Ontario N6A 5B7, Canada\\
  Department of Mathematics and Computer Science,\\ Algoma University, Sault Ste. Marie,\\ Ontario P6A 2G4, Canada \\
dgmckeo2@uwo.ca}

\author{F.\ T.~Brandt\thanks{\textit{fbrandt@usp.br}} , J.~Frenkel\thanks{\textit{jfrenkel@if.usp.br}} ,  S.~Martins-Filho\thanks{Corresponding author, now at Instituto de F\'isica Te\'orica, Universidade Estadual Paulista (UNESP),
Rua Dr. Bento Teobaldo Ferraz 271, Bloco II, Barra Funda,
S\~ao Paulo, S\~ao Paulo 01140-070, Brazil; \textit{s.martins-filho@unesp.br}}}

\address{Instituto de F\'{\i}sica, Universidade de S\~ao Paulo,\\
   S\~ao Paulo, S\~ao Paulo 05508-090, Brazil 
}

\maketitle

\begin{abstract}
A Lagrange multiplier field  restricts the quantum corrections to the Einstein–Hilbert  action at one-loop order, yielding a model that is renormalizable and unitary while reproducing the Einstein field equations in the classical limit.
\end{abstract}

\vspace{1cm}

Quantum mechanics and general relativity are the cornerstones of our  current understanding of physics --- yet these theories do not appear to work well together. In particular, the non-renormalizability of general relativity at high-energies seems to be inconsistent with local quantum field theory. There have been many attempts to solve this fundamental problem. It has been suggested that by extending the number of physical fields through supersymmetric coupling (supergravity) \cite{new1}, extending the concept of particles to strings \cite{new2}, adjusting the canonical structure of the theory (loop gravity) \cite{new3}, invoking non-perturbative behavior (asymptotic safety) \cite{new4,new5}, or by including extra terms into the basic Einstein-Hilbert action \cite{new6,new7}, one may obtain a realistic theory of quantum gravity. Although much progress has been made along these lines, these approaches have not yet proven to be entirely satisfactory.

When the Faddeev-Popov (FP) quantization procedure is applied to the
Einstein-Hilbert (EH) action, then at one-loop order the on-shell
condition eliminates ultraviolet divergences \cite{1,2,3}.
Unfortunately, renormalizability is lost beyond one-loop order, \cite{4,5} or if matter fields are coupled to the graviton \cite{1,6,7,new14}.

For some time, it has been known that by using a
Lagrange multiplier (LM) field $ \lambda$ to ensure that the classical
equations of motion are satisfied, all radiative effects beyond
one-loop order are eliminated \cite{14}.  This suggests that a
LM field used in conjunction with the EH action can result in
the action being consistent with renormalizability
and unitarity while retaining standard General Relativity (GR)
in the classical limit, and at the same time allowing the
standard bosonic matter fields to couple to the metric.
A series of papers develops this approach \cite{15,16,17,18,19}
and in this essay we will outline the steps used.

By considering a simple real integral, the essence of this
approach to the path integral (PI) used to quantize this model
is illustrated.  
If \(L(g)\) is a sufficiently regular function of the $N$ real variables \(g_i\), with isolated non-degenerate stationary points, we find by integrating first over $ \lambda$, then using the resulting $\delta$-function to integrate over $g_i$, that the integral 
\begin{equation}
I \equiv
\int \frac{\mathrm d^N g_i\,\mathrm d^N \lambda_i}{(2\pi)^N}
     \left[\det\left(\frac{\partial^2 L(g_k)}{\partial g_i\,\partial g_j}\right)\right]^{1/2}
     \exp i\left[L(g_k) + \lambda_i
                  \frac{\partial L(g_k)}{\partial g_i}\right],
\label{eq:PI-start}
\end{equation}
can be written as 
\begin{equation}
I \;=\;
\sum_A
\frac{\exp\bigl[i\,L\!\bigl(g_k^{A}\bigr)\bigr]}
     {\det^{1/2}\!\bigl[\partial^2 L\!\bigl(g_k^{A}\bigr)/
                        (\partial g_i\,\partial g_j)\bigr]},
\label{eq:saddle-sum}
\end{equation}
where $g_k^{A}$ satisfies the condition
\begin{equation}
\frac{\partial L\!\bigl(g_k^{A}\bigr)}{\partial g_i}\;=\;0 .
\label{eq:EL-eq}
\end{equation}

If the finite-dimensional integral in Eq. \eqref{eq:PI-start} is generalized to a path integral (PI),
with $g_k$ becoming a field, $\lambda_k$ a
LM 
field, $L(g_i)$ the action for $g_i$, and
$\partial L/\partial g_i=0$ the classical equations of motion, then in
Eq.~\eqref{eq:saddle-sum} 
the exponential collects all
tree-level (classical) diagrams, while the functional determinant
encodes the sum of all one-loop diagrams, these being in the presence of a background field
$g_k^{A}$ that solves the classical equation of motion
\eqref{eq:EL-eq}.  No further contributions to $I$ arise that 
correspond to diagrams beyond one-loop order.
The role of the factor
$\det^{1/2}\bigl[\partial^{2}L(g_k)/(\partial g_i\,\partial
g_k)\bigr]$ in Eq.~\eqref{eq:PI-start} is to ensure that the form of
$I$ is invariant under the change of variables
$g_k' = g_k'(g_j)$\cite{20,21}.

Sources $j_i$ and $J_i$ are now introduced for $g_i$ and $\lambda_i$,
respectively, and we define the \emph{generating functional}
\begin{align}
Z(j_i,J_i)=
\int &{\mathcal D g_i\,\mathcal D\lambda_i}
      \biggl[\det\biggl(
        \frac{\partial^{2}\mathcal L_{\mathrm{cl}}(g_k)}
             {\partial g_i\,\partial g_j}
      \biggr)\biggr]^{\!1/2} \nonumber\\[4pt]
&\times\exp\Biggl\{
      i\!\int d^n x\,
      \Bigl[
        \mathcal L_{\mathrm{cl}}(g_k)
        +\lambda_i\,\frac{\partial\mathcal L_{\mathrm{cl}}(g_k)}{\partial g_i}
        +j_i g_i+J_i\lambda_i
      \Bigr]
      \Biggr\}.
\label{eq:Z-def}
\end{align}
Following Ref.~\refcite{22}, we introduce the background classical fields
\begin{equation}
\bar g_i \;\equiv\; \frac{\partial W}{\partial j_i}, 
\qquad
\bar\lambda_i \;\equiv\; \frac{\partial W}{\partial J_i},
\label{eq:bg-fields}
\end{equation}
where \(W(j_i,J_i)=-i\ln Z(j_i,J_i)\). We then perform a Legendre
transformation from \(j_i,J_i\) to \(\bar g_i,\bar\lambda_i\), so that
\(\Gamma[\bar g(x),\bar\lambda(x)]\) becomes the generating functional
of one-particle-irreducible diagrams. Thus, we have that
\begin{equation}
\allowdisplaybreaks
\label{eq:Gamma-field}
\begin{aligned}
&e^{i\Gamma[\bar g,\bar\lambda]} =
\int\!\mathcal{D}h_i\,\mathcal{D}H_i\;
      \det\nolimits^{1/2} \Bigl(
        \frac{\partial^{2}\mathcal L_{\text{cl}}}{\partial h_i\,\partial h_j}
      \Bigr)
\\
&\times
\exp\Bigl\{\,i\!\int d^{n}x\,
      \bigl[
        \mathcal L_{\text{cl}}(\bar g_i+h_i)
        +(\bar\lambda_i+H_i)
          \partial_{h_i}\mathcal L_{\text{cl}}(\bar g_i+h_i)
        +j_i h_i+J_i H_i
      \bigr]\Bigr\},
\end{aligned}
\end{equation}
where now $j_i =  - \delta \Gamma / \delta \bar g_i$   and $J_i = -\delta \Gamma/ \delta \bar \lambda_i$.
The quantum fluctuations around the background fields $\bar g_i$ and $\bar\lambda_i$ are represented by $h_i$ and $H_i$, respectively.

If \(\int d^{n}x\,\mathcal L_{\text{cl}}\bigl(g^{i}\bigr)\) is invariant under the gauge transformations
\begin{equation}
g_i \to g_i + R_{ij}(g)\,\xi^{j},
\label{eq:gauge}
\end{equation}
then the action
\(
\int d^{n}x\,[\mathcal L_{\text{cl}} + \lambda_i\,\partial_{g_i}\mathcal L_{\text{cl}}]
\)
is invariant under the transformation  \eqref{eq:gauge} as well as
\begin{subequations}\label{eq:lambdagauge}
\begin{align}
\lambda_i &\to \lambda_i + R_{ij}(g)\,\zeta^{j},
\label{eq:lambda-gauge-a}\\[2pt]
\lambda_i &\to \lambda_i
            + \lambda_m\,
              \frac{\partial R_{ij}(g)}{\partial g_m}\,\xi^{j}.
\label{eq:lambda-gauge-b}
\end{align}
\end{subequations}
Applying the  FP quantization procedure adds to the
exponent in Eq.~\eqref{eq:Gamma-field} a gauge-fixing term
together with the usual
FP-ghost contribution.

If the gauge fixing
is chosen to preserve background gauge invariance
[i.e.\ Eqs.~\eqref{eq:gauge}-\eqref{eq:lambdagauge} with
$g_{i}\!\to\!\bar g_{i}$,\,
$\lambda_{i}\!\to\!\bar\lambda_{i}$], the generating functional becomes
\begin{align}
\allowdisplaybreaks
&\exp\bigl[i\Gamma(\bar g_i,\bar\lambda_i)\bigr]
  = \int\!\mathcal D h_i\,\mathcal D H_i\;
     \det\nolimits^{1/2}    \frac{\partial^{2}}
            {\partial h_i\partial h_j}\left(
       \mathcal L_{\mathrm{cl}}(\bar g+h)
       -\frac{1}{\alpha}\,
        F_{ik}(\bar g)h_k F_{i\ell}(\bar g)\,h_\ell
     \right) \nonumber\\
  &\times
     \det \bigl[F_{ij}(\bar g)\,R_{jk}(\bar g+h)\bigr]\;
     \exp\Biggl\{
       i\!\int d^{n}x\,
       \biggl[
         \mathcal L_{\mathrm{cl}}(\bar g+h) 
         +\bigl(\bar\lambda_i+H_i\bigr)\,
            \left .\frac{\partial\mathcal L_{\mathrm{cl}}(g)}{\partial g_i}\right|_{g=\bar{g}+h}
        \nonumber\\ &-\frac{1}{2\alpha}\bigl[F_{ij}(\bar g)\,h_j\bigr]^2 
         -\frac{1}{\alpha}\bigl[F_{ij}(\bar g)\,h_j\bigr]
                         \bigl[F_{i\ell}(\bar g)\,H_\ell\bigr]
         +j_i h_i+J_i H_i
       \biggr]
     \Biggr\}.
\label{eq:Gamma-bg}
\end{align}
The gauge fixing term to the exponential involves the introduction of a gauge parameter $\alpha$.

Bosonic matter fields $\Phi_{I}$ can be coupled to the gauge fields
$g_{i}$ through a conventional Lagrangian
$\mathcal L_{M}(\Phi_{I},g_{i})$, provided the gauge
transformation \eqref{eq:gauge} is supplemented by
\(
\Phi_{I} \to \Phi_{I}+R_{I j}(\Phi)\,\xi^{j}
\),
which leaves $\int d^{n}x\,\mathcal L_{M}(\Phi_{I},g_{i})$ invariant.
We do not introduce an LM field to restrict the radiative effects of
$\Phi_{I}$ to one-loop order, as that would generate Landau-pole
singularities in the running gauge coupling \cite{14}.

We now outline two approaches to the radiative corrections implied by
Eq.~\eqref{eq:Gamma-bg}.  In the first approach, the integral over
$H_{i}$ is performed first, leaving a functional
$\delta$-function,
$\mathop{\delta}\bigl(
  {\partial \mathcal L_{\text{cl}}(\bar g+h)}/{\partial h_{i}}
  -{1}/{\alpha}\,F_{ij}(\bar g) F_{jk}(\bar g)\,h_{k}
\bigr)$,
after which the sources $j_{i}$ and $J_{i}$ are set to zero and the integration over $h_{i}$ is carried out.  
This leads to the closed form \cite{14,15,16,17,18,19}
\begin{equation}
\begin{aligned}
&\exp\bigl[i\,\Gamma(\bar g_i,0)\bigr]
  = \sum_{A} \det\nolimits^{-1/2}  \frac{\partial^{2}}
             {\partial h_i\,\partial h_j}\left(
      \mathcal L_{\text{cl}}(\bar g+h^{A})
        -\frac{1}{\alpha}F_{ik}(\bar g)\,h_{k}^{A}
        F_{i\ell}(\bar g)\,h_{\ell}^{A}
      \right)\;
\\
  &\times     \det\bigl[F_{ij}(\bar g)\,R_{jk}(\bar g+h^{A})\bigr]
     \exp \left\{\,i\!\int d^{n}x\,
       \Bigl[
         \mathcal L_{\text{cl}}(\bar g+h^{A})
         -\frac{1}{2\alpha}\bigl(F_{ij}(\bar g)\,h_{j}^{A}\bigr)^{2}
       \Bigr]\right\},
\end{aligned}
\label{eq:Gamma-hA}
\end{equation}
where $h_i^A$ comes from the $\delta$-function.
The background field $\bar g_i$ need not satisfy the classical equation of motion.

\begin{figure}[ht]
  \centering
  \includegraphics[width=.9\textwidth]{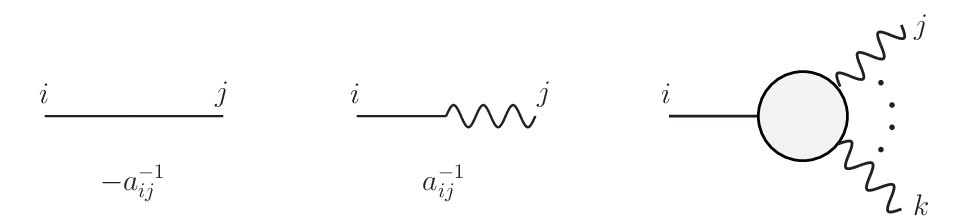}
\caption{Propagators and vertices. The fields \(h_i\) and \(H_i\) are represented by wavy and solid lines, respectively.}
  \label{fig:prop-vert}
\end{figure}

\begin{figure}[h]
  \centering
  \includegraphics[width=0.4\textwidth]{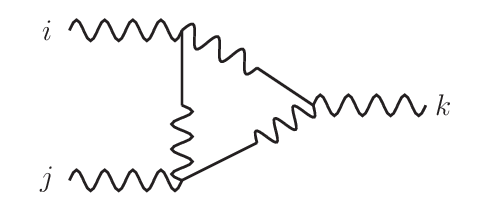}
\caption{A representative one-loop diagram. In this case, the diagram contributes to the three-point function \(h_i h_j h_k\).}
  \label{fig:one-loop}
\end{figure}

A second approach to Eq.~\eqref{eq:Gamma-bg} is to derive the Feynman
rules for the fields \(h_i\) and \(H_i\) and analyze the resulting
diagrams.  
The terms bilinear in these fields can be written in the form
\begin{equation}
\frac{1}{2}\begin{pmatrix} h_i & H_i \end{pmatrix}
\begin{pmatrix} a_{ij} & a_{ij} \\ a_{ij} & 0 \end{pmatrix}
\begin{pmatrix} h_j \\ H_j \end{pmatrix}.
\end{equation}
Inverting this matrix, we obtain the propagators
\begin{equation}
\begin{pmatrix} a_{ij} & a_{ij} \\ a_{ij} & 0 \end{pmatrix}^{-1}
=
\begin{pmatrix}
     0          & a^{-1}_{ij}\\
     a^{-1}_{ij} & -a^{-1}_{ij}
\end{pmatrix}.
\label{eq:bilinear-inv}
\end{equation}
Thus
the propagator for the field \(h_i\) vanishes, the propagator of the LM field
\(H_i\) carries the ``wrong sign'',
and
the mixed propagator for \(h_i\)-\(H_i\) is simply \(a^{-1}_{ij}\), the
same as would occur for \(h_i\) alone if \(H_i\) had not been
introduced.  Moreover, \(H_i\) enters \emph{at most linearly} in any
vertex.
These features eliminate all Feynman diagrams beyond one-loop
order and restrict the one-loop diagrams to those with no external LM
legs and only mixed \(h\)-\(H\) propagators (see Fig. \ref{fig:prop-vert} and Fig. \ref{fig:one-loop}).

Combining those one-loop diagrams with the contribution from the first
functional determinant in Eq.~\eqref{eq:Gamma-bg} reproduces exactly
the one-loop diagrams that would arise if no LM field were present.
The second determinant is 
the usual FP-ghost determinant at
one-loop order. No higher-loop contributions can occur.

This diagrammatic approach does not determine the explicit contribution of $h_i^{A}$ in Eq. \eqref{eq:Gamma-hA}; likewise, ordinary Feynman graphs in Yang–Mills theory fail to reveal the occurrence of instanton solutions.
Those classical solutions \(h_i^{A}(x)\) could be said to have ``dark-matter-like''
properties: they couple only through the propagation of \(g_i\) and
\(\Phi_I\) and do not themselves propagate.

We now specialize to the Einstein-Hilbert (EH) Lagrangian.  Replacing
\(\mathcal L_{\text{cl}}\) by \(\mathcal L_{\text{EH}}\) in
Eq.~\eqref{eq:Gamma-bg} gives
\begin{equation}
\int d^{n}x\,[\mathcal L_{\text{EH}}(g)+\lambda_i\partial_{g_i}\mathcal
L_{\text{EH}}(g)]
  \longrightarrow
\frac{1}{\kappa^{2}}\int d^{n}x\,\sqrt{-g}\,
  \bigl[-g^{\mu\nu}R_{\mu\nu}(g)+\lambda^{\mu\nu}
G_{\mu\nu} 
\bigr],
\label{eq:EH-action}
\end{equation}
with \(\kappa^{2}=16\pi G_{N}\), {\mbox{$G_{\mu\nu} =  R_{\mu\nu}(g)-\tfrac12 g_{\mu\nu}R(g)$}} is the Einstein tensor and our sign conventions are such that
\(R^\rho_{~\;\sigma\mu\nu} =
\partial_\mu\,\Gamma^\rho_{\sigma\nu} -  
\partial_\nu\,\Gamma^\rho_{\sigma\mu} +  
\Gamma^\lambda_{\sigma\nu}\,\Gamma^\rho_{\lambda\mu} -  
\Gamma^\lambda_{\sigma\mu}\,\Gamma^\rho_{\lambda\nu}
\).

In the case of the Einstein–Hilbert action, the generic gauge transformation \eqref{eq:gauge} corresponds to infinitesimal diffeomorphisms of the metric, given by
\begin{equation}
\delta g_{\mu\nu}=g_{\rho\nu}\,\partial_{\mu}\xi^{\rho}
                 +g_{\mu\rho}\,\partial_{\nu}\xi^{\rho}
                 +\xi^{\rho}\,\partial_{\rho}g_{\mu\nu}.
\label{eq:diff-gauge}
\end{equation}
If $\bar g_{\mu\nu}$ is the background to the metric,
Refs.~\refcite{1,2,3} show that the one-loop divergences from the EH sector
alone are
\begin{equation}
\frac{1}{16 \pi^2 \varepsilon}\int d^{n}x\,\sqrt{-\bar g}\,
  \Bigl[\frac{1}{120}\,R^{2}(\bar g)+\frac{7}{20}\,
        R_{\mu\nu}(\bar g)R^{\mu\nu}(\bar g)\Bigr],
\label{eq:EH-div}
\end{equation}
with \(\varepsilon=2-\tfrac{n}{2}\).

Since
\begin{equation}
R_{\mu\nu}=G_{\mu\nu}+\frac{1}{2-n}g_{\mu\nu}g_{\lambda\sigma}G^{\lambda\sigma},
\label{eq:Rmunu-A}
\end{equation}
the divergences in Eq.~\eqref{eq:EH-div} can be absorbed into the
background field \(\bar\lambda_{\mu\nu}\) that accompanies
\(\lambda_{\mu\nu}\) in Eq.~\eqref{eq:EH-action}.
Renormalizing the background field $\bar\lambda_{\mu\nu}$ to eliminate
divergences is \emph{distinct} from the usual renormalization procedure
in which the couplings, masses and fields that characterize a theory
absorb divergences.

If the metric $g_{\mu\nu}$ is coupled to \emph{bosonic} matter fields
(either scalar or vector), the one-loop divergences that involve the
propagation of $g_{\mu\nu}$ are again proportional to
$R_{\mu\nu}$\cite{1,6,7} and can therefore be absorbed into
$\bar\lambda_{\mu\nu}$.\footnote{For bosonic matter we introduce \emph{no} LM field, so
  diagrams containing bosonic matter loops occur at all orders.}
The usual matter-field divergences can still be handled in the standard way, i.e.\ by renormalizing the parameters of the matter sector.  Divergences
simply proportional to $\sqrt{-g}$ require renormalization of a
cosmological constant\cite{23}. 

\begin{figure}[h]
  \centering
    \subfloat[\label{fig:3a} Tree-level diagrams.]{\includegraphics[width=0.8\textwidth]{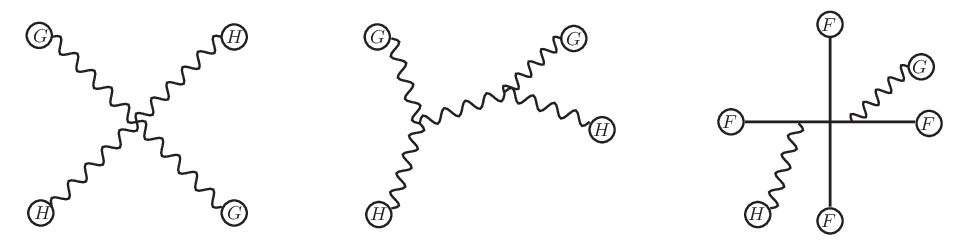}}\\
    \subfloat[\label{fig:3b} One-loop diagrams involving the metric field.]{\includegraphics[scale=0.618]{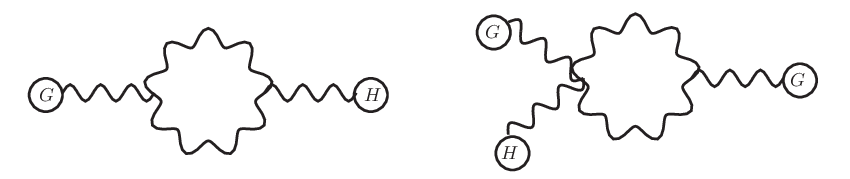}}\\
  \subfloat[\label{fig:3c}  Loop diagrams involving the scalar field.]{\includegraphics[scale=0.618]{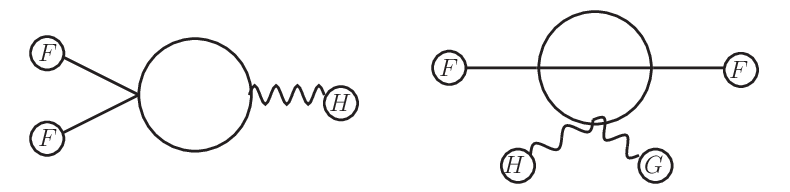}}
  \caption{Typical diagrams contributing to the EH action.  The contributions of the fields   $ \bar{g}_{\mu \nu} $ and $ h_{\mu \nu}^{A} $, $ \bar{\Phi} $ are denoted by    \Circled{\scriptsize $G$}, \Circled{\scriptsize  $H$},  \Circled{\scriptsize  $F$}, respectively.} 
  \label{fig:scalar-diags}
\end{figure}

Applying the saddle-point approach to
Eq.~\eqref{eq:Gamma-bg} with the EH action supplemented by the action for a scalar field with a
quartic self-interaction, we obtain typical contributing diagrams shown
in Fig.\,\ref{fig:scalar-diags}.

If \emph{fermionic} matter fields are present they couple to gravity via
the Einstein-Cartan (EC) action\cite{24,25}.  Quantizing this theory
requires handling both diffeomorphism and local Lorentz invariance; see Refs.~\refcite{26,27} for details.

The unitarity of the model defined by Eq. \eqref{eq:Gamma-field} can be demonstrated by applying the Cutkosky cutting rules, which relate the imaginary part of the amplitude at some loop order to the total  cross sections
at one lower order \cite{19}.

\begin{figure}[h]
  \centering
    \includegraphics[width=0.6\textwidth]{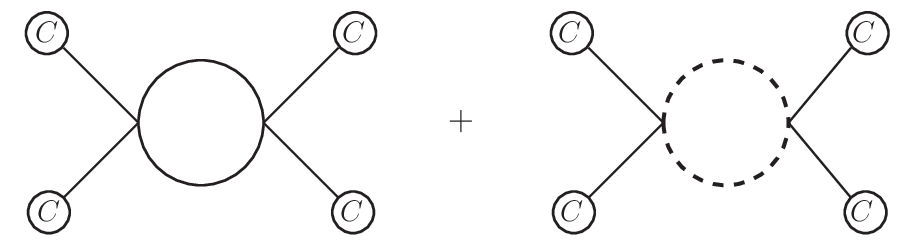}
  \caption{One-loop four-point diagrams for the
  \(C\) field. Solid lines denotes the field $C$, dashed lines represent the field $B$.}
  \label{fig:four-point-C}
\end{figure}

\begin{figure}[h]
  \centering
   \includegraphics[width=0.65\textwidth]{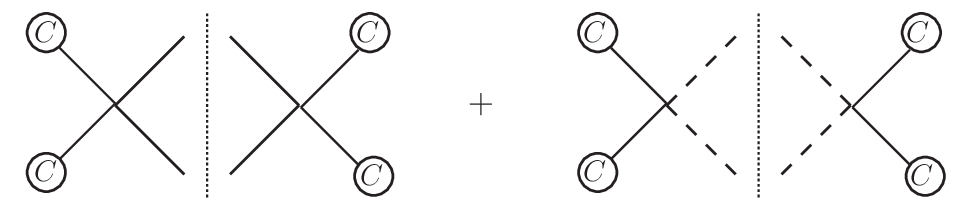}
  \caption{Cut diagrams illustrating the
tree-level cross section for the \(C\) field.}
  \label{fig:cut-diagrams-C}
\end{figure}

To illustrate unitarity when an LM field is present,  
consider a massive real scalar field \(A\) with a quartic self-interaction that
also couples to an LM field \(B\), so that \cite{18}
\begin{equation}
\mathcal L_{\mathrm{cl}}(A)+B\, \frac{\partial \mathcal L_{\mathrm{cl}}}{\partial A}(A)=
  -\frac12 A(\Box+m^{2})A
  -\frac{g}{4!}\,A^{4} - B\left(
  \Box A + m^2 A + \frac{g}{3!} A^3
  \right) .
\label{eq:LM-scalar}
\end{equation}
Replacing \(A\) by the shifted field \(C=A+B\) diagonalizes the free
propagators for \(B\) and \(C\).  
Although the interaction vertices are now more complicated,
\emph{only one-loop diagrams} arise even when the theory is expressed in
terms of the LM field
\(B\) and the physical field \(C\).
Explicit calculation shows, for example, that the four-point amplitude for
\(C\)-field scattering at one loop obeys the usual Cutkosky cutting rules for
the forward-scattering amplitude
(see Figs. \ref{fig:four-point-C} and \ref{fig:cut-diagrams-C}). In Ref.~\refcite{18}, we have also shown that this can be extended for the contributions coming from the factor $\det^{1/2}\bigl[\partial^{2} \mathcal L_{\text{cl}}(A)/\partial A^2 \bigr]$, which is required to a consistent quantization of this model when the LM field is present.
Furthermore, the path integral in 
Eq.~\eqref{eq:Gamma-bg}  is BRST-invariant, ensuring unitarity to all orders.

It has long been known that EH gravity is one-loop renormalizable only
\emph{on shell}\cite{1,2,3}.  This requirement is absent in QED or
Yang-Mills, which are renormalizable off shell\cite{28}.  In our
approach, the LM field 
couples to the classical
equations of motion, so 
divergences 
are eliminated by absorbing them into the LM field leaving the
gravitational coupling $\kappa^{2}$ untouched.  Consequently $\kappa^{2}$
does not ``run'' with the renormalization scale
$\mu^2$, unlike the QED or Yang-Mills
coupling.
The LM field itself, and the parameters of the matter sector and the
cosmological constant, \emph{do} depend on the scale $\mu^2$ and vary as
$\mu^2$ changes.

Applying the same LM mechanism to Einstein-Cartan gravity should likewise suppress multi-loop graphs, yielding a renormalizable, unitary coupling of fermions to gravitons and bringing gravity into the Standard-Model fold. This issue is currently being investigated.

\section*{Acknowledgments}
We are grateful to T. N. Sherry for earlier discussions that helped motivate this work.
F.\ T.\ Brandt and J.\ F.\ thank CNPq (Brazil) for financial support.  S.\ M.-F.\ thanks FAPESP for financial support. 
This study was financed, in part, by the São Paulo Research Foundation (FAPESP), Brasil. Process Number \#2025/16156-7.

\section*{ORCID}
D. G. C. McKeon\,\orcidlink{0000-0002-6152-4495}
\url{https://orcid.org/0000-0002-6152-4495}\\
F. T. Brandt\,\orcidlink{0000-0001-7873-7684}
\url{https://orcid.org/0000-0001-7873-7684}\\
J. Frenkel\,\orcidlink{0000-0003-1178-1001}
\url{https://orcid.org/0000-0003-1178-1001}\\
S. Martins-Filho\,\orcidlink{0000-0003-4195-2713}
\url{https://orcid.org/0000-0003-4195-2713}.


\begin{thebibliography}{99}
\bibitem{new1} D.Z. Freedman and A. Van Proeyen, ``Supergravity,'' Cambridge University Press, Cambridge, UK (2012).

\bibitem{new2} P. West, ``Introduction to Strings and Branes,'' Cambridge University Press, Cambridge, UK (2012).

\bibitem{new3} R. Gambini and J. Pullin, ``A First Course in Loop Quantum Gravity,'' Oxford University Press, Oxford, UK (2011).

\bibitem{new4} S. Nagy, Ann. Phys. \textbf{350}, 310 (2014).

\bibitem{new5} I.B. Khriplovich, Yad. Fiz. \textbf{10} (1969) 409 [Sov. J. Nucl. Phys. \textbf{10} (1970) 23].

\bibitem{new6} K. Stelle, Phys. Rev. D \textbf{16}, 953 (1977).

\bibitem{new7} P.D. Mannheim, Found. Phys. \textbf{42}, 388 (2012).

\bibitem{1} G.\,'t Hooft and M.\,Veltman, \emph{Ann.\ Inst.\ H.\ Poincaré} -- Les Houches 1975,
            \textbf{20} (1974) 69.
\bibitem{2} M.\,Veltman in: R.\,Balian and J.\,Zinn-Justin (Eds.),
            \emph{Methods in Field Theory}, North-Holland, Amsterdam (1976).
\bibitem{3} G.\,'t Hooft, ``Perturbative Quantum Gravity'', Erice 2002,
            \url{http://www.staff.science.uu.nl/~hooft101/lectures/erice02.pdf}.
\bibitem{4} M.\,H.\,Goroff and A.\,Sagnotti, \emph{Nucl.\ Phys.\ B}
            \textbf{266} (1986) 709.
\bibitem{5} A.\,E.\,M.\,van de Ven, \emph{Nucl.\ Phys.\ B} \textbf{378}
            (1992) 309.
\bibitem{6} S.\,Deser, H.\,S.\,Tsao and P.\,van Nieuwenhuizen,
            \emph{Phys.\ Rev.\ D} \textbf{10} (1974) 3337.
\bibitem{7} S.\,Deser and P.\,van Nieuwenhuizen, \emph{Phys.\ Rev.\ D}
            \textbf{10} (1974) 411.

\bibitem{new14}
J.~Frenkel and S.~Martins-Filho,
\emph{Phys. Rev. D} \textbf{113}, no.8, 085012 (2026).

\bibitem{14} D.\,G.\,C.~McKeon and T.\,N.~Sherry,
             \emph{Can.\ J.\ Phys.} \textbf{70} (1992) 441.

\bibitem{15} F.\,T.~Brandt, J.~Frenkel and D.\,G.\,C.~McKeon,
             \emph{Can.\ J.\ Phys.} \textbf{98} (2020) 344.

\bibitem{16} F.\,T.~Brandt, J.~Frenkel, D.\,G.\,C.~McKeon and G.\,S.\,S.~Sakoda,
             \emph{Phys.\ Rev.\ D} \textbf{100} (2019) 125014.

\bibitem{17} F.\,T.~Brandt, J.~Frenkel, D.\,G.\,C.~McKeon and S.~Martins-Filho,
             \emph{Ann.\ Phys.} \textbf{427} (2021) 168426.

\bibitem{18} F.\,T.~Brandt, J.~Frenkel, D.\,G.\,C.~McKeon and S.~Martins-Filho,
             \emph{Ann.\ Phys.} \textbf{434} (2021) 168659.

\bibitem{19} F.\,T.~Brandt, J.~Frenkel, D.\,G.\,C.~McKeon and S.~Martins-Filho,
             \emph{Ann.\ Phys.} \textbf{480} (2025) 170101.

\bibitem{20} F.\,T.~Brandt and S.~Martins-Filho,
             \emph{Ann.\ Phys.} \textbf{453} (2023) 169323.

\bibitem{21} F.\,T.~Brandt, S.~Martins-Filho and D.\,C.\,C.~McKeon,
             \emph{Eur.\ Phys.\ J.\ C} \textbf{84} (2024) 399.

\bibitem{22} L.\,F.~Abbott,
             \emph{Nucl.\ Phys.\ B} \textbf{185} (1981) 189;
             \emph{Acta Phys.\ Pol.\ B} \textbf{13} (1982) 33.

\bibitem{23} J.~Buchbinder and I.\,L.~Shapiro,
             \emph{Introduction to Quantum Field Theory with Applications
             to Quantum Gravity}, Oxford University Press, Oxford UK (2021).

\bibitem{24} T.\,W.\,B.~Kibble,
             \emph{J.\ Math.\ Phys.} \textbf{2} (1961) 212.

\bibitem{25} F.\,W.~Hehl, P.~von der Heyde, G.\,D.~Kerlick and J.\,M.~Nester,
             \emph{Rev.\ Mod.\ Phys.} \textbf{48} (1976) 393.

\bibitem{26} F.\,T.~Brandt, J.~Frenkel, S.~Martins-Filho and D.\,G.\,C.~McKeon,
             \emph{Ann.\ Phys.} \textbf{462} (2024) 169607.

\bibitem{27} F.\,T.~Brandt, J.~Frenkel, S.~Martins-Filho and D.\,G.\,C.~McKeon,
             \emph{Ann.\ Phys.} \textbf{470} (2024) 169801.

\bibitem{28} S.~Weinberg,
             \emph{The Quantum Theory of Fields}, vols.\ I \& II,
             Cambridge University Press, Cambridge UK (1996).

\end{thebibliography}
\end{document}